\documentclass[12pt,twoside,a4paper]{article}
\usepackage[hmargin=25mm,vmargin=20mm]{geometry}
\usepackage{epigraph}
\usepackage[T1]{fontenc}
\usepackage{csquotes}

\usepackage{blindtext}
\usepackage[pdfpagelabels=false,pdfborder={0 0 0}]{hyperref}

\usepackage{amsmath}
\usepackage{graphicx}
\usepackage{listings}
\usepackage{textcomp}
\usepackage{ amssymb }

\newcommand{\centercode}[1]{ \vspace{2mm}  \noindent\texttt{#1}
\vspace{2mm}}

\title{Formalizing May's Theorem}
\author{Li, Kwing Hei}
\date{}
\begin{document}
\maketitle

\begin{abstract}

This report presents a formalization of May's theorem in the proof assistant Coq. It  describes how the theorem statement is first translated into Coq definitions, and  how it is subsequently proved. Various aspects of the proof and related work are discussed. To the best of the author's knowledge, this project is the first documented attempt in mechanizing May's Theorem.
\end{abstract}

\section{Introduction}


In 1952, Kenneth May published a mathematical theorem on social choice theory, which establishes a set of necessary and sufficient conditions for simple majority voting~\cite{mays}. This result is now known as \textbf{May's theorem}. Though not as famous as Arrow's theorem~\cite{arrows} and the Gibbard-Satterthwaite theorem~\cite{gibbard}, May's theorem  is still considered to be one of the "minor classics" in voting theory~\cite{barry}. While the other above theorems in economics have been formalized in proof assistants~\cite{formalize_arrow_isabelle}\cite{formalize_arrow_mizar}, May's theorem has never been mechanically proved. 

In this report, I present the first mechanized proof of May's theorem, which I implemented in \textbf{Coq}~\cite{coq}. Proving the theorem in Coq not only increments the library of all proved theorems, but also provides insights in mechanizing  results from social choice theory in a type theory based proof assistant.  The structure of the Coq proof differs  from the conventional proof sketch, and   this report explains various subtle details of it.

In this document, I first present the high-level statement of the theorem in Section~\ref{mays}. In Section~\ref{definitions}, I then elaborate how this statement is  translated into Coq definitions. Section~\ref{if} and \ref{onlyif} then detail the structure of the Coq proof in the if direction and only if direction, respectively. In Section~\ref{discussion}, I consider the nature of the proof itself by discussing its usefulness, correctness, and the extra extensionality axiom the proof relies on. Section~\ref{conclusion} concludes the report and examines various future directions.

It should be noted that this document only mentions a small but important subset of the  lemmas that constitute the main backbone of the Coq proof. A large number of relatively minor lemmas that the proof uses are omitted from this report for the sake of brevity. Although many of these lemmas appear straightforward at first glance, their mechanical proofs are quite convoluted and require much case analysis or extensive proof by induction. One example is the following lemma which states if two distinct elements, say $x$ and $y$, are members of a list containing no duplicates, the list can be expressed as the concatenation of five lists, two of which are singleton lists containing $x$ and $y$, respectively\footnote{The Coq keyword \texttt{Lemma} allows us to write a proposition for which its proof is then built using tactics.}\footnote{\texttt{NoDup l} expresses that the list \texttt{l} contains no duplicates. }:

\centercode{Lemma two\_elements\_exist \{A:Type\} (l:list A) x y \`{}\{!NoDup l\} \`{}\{x$\neq$y\}: \\ \indent In x l → In y l →  $\exists $l1 l2 l3, \\ \indent l=l1++[x]++l2++[y]++l3 $\lor$ l=l1++[y]++l2++[x]++l3.}

\section{May's Theorem\label{mays}}
We now establish certain definitions following social choice theory nomenclature. 

Consider an election with exactly two candidates $x$ and $y$, and a finite set of voters. Each voter can cast a vote, called a \textbf{preference}, indicating that they either prefer $x$ over $y$, $y$ over $x$, or that they are indifferent between $x$ and $y$. 

A \textbf{social choice function} is a function that maps each possible list of preferences to a unique result, either $x$ wins, $y$ wins, or there is a tie between the two candidates.

Not all social functions are what we would perceive to be fair. May identifies several properties.

A social choice function is \textbf{anonymous} if it is a symmetric function of its arguments. Each voter is treated the same by the social choice function, and swapping the preference of any two voters will yield the same election result.

A social choice function is \textbf{neutral} if flipping the preference of each voter will also flip the function's result. (Here, an indifference preference remains unchanged after a flip). Both candidates are treated equally by the social choice function and swapping the names of the candidates will not affect the final result.

A social choice function is \textbf{monotone} if for any list of preferences where the result is indifferent or favorable to $x$, changing any voter's preference in a positive way towards $x$ results in $x$ winning the election. (Here, changing a preference in a positive way towards $x$ means changing a $y$ vote to an indifference one, a $y$ to a $x$, or an indifference vote to a $x$.) This means the social choice function responds to changes of individual preferences in a "positive" manner.

Suppose the number of people who vote for $x$ is $a$ and the number of people who vote for $y$ is $b$. The \textbf{simple majority function} is the social choice function that decides  $x$ wins if $a>b$, $y$ wins if $b>a$, or that the two candidates tie otherwise. 

May's theorem states that a social choice function is anonymous, neutral, and monotone if and only if it is the simple majority function. This means that anonymity, neutrality, and montonicity form a set of necessary and sufficient conditions for a social choice function to be the simple majority function.

\section{Definitions and assumptions in Coq\label{definitions}} 

In my proof, I used the \texttt{Finite} typeclass, from the library coq-std++~\cite{stdpp}, to express the finite set of voters being considered\footnote{The Coq keyword \texttt{Context} declares variables in the context of a section. In this case, the type \texttt{voter} belonging to the typeclass \texttt{Finite}.}:

\centercode{Context \`{}\{Finite voter\}.}

As shown below, this typeclass\footnote{Typeclasses are defined with the Coq keyword \texttt{Record}.} provides three fields for the \texttt{voter} type: the field \texttt{enum voter} is a list containing all the voters, \texttt{NoDup\_enum voter} is a proposition stating that the list contains no duplicates, and \texttt{elem\_of\_enum voter} is a proposition stating that every term of the type \texttt{voter} is an element of the list. All three fields turn out to be equally important and are used extensively throughout the mechanical proof.

\centercode{Record Finite (A : Type) (EqDecision0 :\ EqDecision A) : Type := \\
\indent Build\_Finite \{ enum :\ list A;\\
\indent \indent	NoDup\_enum :\ NoDup enum;\\
  \indent \indent elem\_of\_enum : $\forall$ x :\ A, x $\in$ enum \}.}

Because whenever we want to reason about the two candidates (whether it is the preference of a voter or the result of the social choice function), we always have a third case where they tie, it is convenient to express the \texttt{candidate} type as \texttt{option bool}. Assuming the candidates are $x$ and $y$, this allows us to represent $x$ being preferred over $y$ by \texttt{Some true}, $y$ being preferred by \texttt{Some false}, and a tie with \texttt{None}. The use of \texttt{bool} also allows us to reason about the duality of preferences with simple \texttt{bool} functions like \texttt{negb}. The type for \texttt{preferences} and \texttt{social\_choice\_function} are   simple function types\footnote{The Coq keyword \texttt{Definition} binds a term to a variable name.}\footnote{The text description of the theorem talks about a list of preferences, but in Coq, it is simpler to model the preferences of voters as a function type.}:

\centercode{Definition preferences := voter → candidate.\\
Definition social\_choice\_function := preferences → candidate.
}

The \texttt{anonymous} property is expressed via a \texttt{swap} function, which swaps the preferences of two voters, while leaving the rest unchanged:

\centercode{Definition swap (v1 v2:voter) (p:preferences) :=\\
 \indent   fun v => \\
 \indent \indent if bool\_decide (v=v1) then p v2 \\
 \indent \indent else if bool\_decide (v=v2) then p v1 \\
 \indent \indent else p v.\\
Definition anonymous (scf:social\_choice\_function):=\\
 \indent   $\forall$ p v1 v2, scf p = scf (swap  v1 v2 p).
}

The \texttt{neutral} property is expressed via a \texttt{flip} and \texttt{flip\_vote} function. The former flips a single \texttt{candidate} term, while the latter expresses the flipped version of the preferences of all voters: 

\centercode{Definition flip cand := \\
 \indent   match cand with \\
   \indent  | Some b => Some (negb b)\\
 \indent    | None => None\\
  \indent   end.\\
Definition flip\_vote (p:preferences) :=\\
 \indent    fun v:voter=> flip (p v).\\
Definition neutral (scf:social\_choice\_function) :=\\
  \indent   $\forall$ p, scf p =  flip (scf (flip\_vote p)).
}

The \texttt{monotone} property is expressed with an \texttt{update} function that updates the preference of a single voter while leaving the rest unchanged:

\centercode{Definition update v i (p:preferences) := \\
\indent fun v':voter => \\
\indent \indent if bool\_decide(v'=v) then i else p v'.\\
Definition monotone (scf:social\_choice\_function) :=\\
\indent    $\forall$ p , (scf p = Some true $\lor$ scf p = None)→  \\
\indent \indent     ($\forall$ v, p v = Some false → scf (update v None p) = Some true) $\land$ \\
\indent \indent     ($\forall$ v, p v = Some false → scf (update v (Some true) p) = Some true) $\land$ \\
\indent  \indent    ($\forall$ v, p v = None → scf (update v (Some true) p) = Some true).}

After defining  \texttt{count}, \texttt{count\_helper}, and various predicate functions for candidates like \texttt{is\_some\_true}, we can express the \texttt{majority\_election} rule as such:

\centercode{Definition majority\_election (p:preferences) :=\\
\indent    let true\_num := count is\_some\_true p in \\
 \indent     let false\_num := count is\_some\_false p in\\
 \indent     if bool\_decide (false\_num < true\_num ) then Some true \\
 \indent  else if bool\_decide (true\_num < false\_num) then Some false \\
 \indent  else None.}

Finally, one can express the main \texttt{mays\_thm} with the following proposition\footnote{The Coq keyword \texttt{Theorem} works the same as \texttt{Lemma}. }:

\centercode{Theorem mays\_thm scf: \\
\indent anonymous scf $\land$ neutral scf $\land$ monotone scf $\leftrightarrow$  scf  = majority\_election .}

One must also mention that the mechanical proof requires an additional axiom that is outside Coq's usual type-based calculus\footnote{The Coq keyword \texttt{Axiom} extends the environment with an axiom.  }. Specifically, we include the axiom of  functional extensionality. The reason for this inclusion is discussed in subsection~\ref{extensionality}. 

\centercode{Axiom functional\_extensionality \{A B\} (f g :\ A → B):\\
\indent  ($\forall$ x, f x = g x) → f = g.
}

\section{If direction\label{if}}

For the if direction of the theorem, we prove that the simple majority voting system is anonymous, neutral, and monotone.

\subsection{Anonymity}

The anonymous property of the simple majority function is mainly supported by the following invariant on \texttt{swap}\footnote{The argument \texttt{f} ranges over predicate functions for candidates, like \texttt{is\_some\_false}.}:

\centercode{Lemma swap\_invariant\_count p f l v1 v2 \`{}\{!NoDup l\}:\\
\indent In v1 l → In v2 l → count\_helper f p l = count\_helper f (swap v1 v2 p) l.
}

This lemma states that given a subset of voters, swapping the preferences of any two voters within the subset does not change the overall number of each type of preference. In fact this lemma relies on another similar one, which asserts that swapping the preferences of any two voters \textbf{not} within the subset also does not change the frequency of the types of preferences:

\centercode{Lemma swap\_not\_in\_list p f v1 v2 l:\\  \indent $\thicksim$In v1 l → $\thicksim$In v2 l → count\_helper f p l =\\ \indent \indent count\_helper f (swap v1 v2 p) l.
}

It is also possible that we choose to swap the preference of a voter with itself, in which case, the list of preferences remains unchanged: 

\centercode{Lemma swap\_same v  p:\ (swap v v p) = p.}

\subsection{Neutrality\label{neutrality}}

The neutrality of the simple majority function is proved by showing that the frequency of \texttt{Some true} and \texttt{Some false} are swapped whenever we perform a \texttt{flip\_vote} on a list of preferences:

\centercode{Lemma flip\_reverse\_count1 p l:\\ \indent count\_helper is\_some\_false (flip\_vote p) l = \\
\indent count\_helper is\_some\_true p l.\\
Lemma flip\_reverse\_count2 p l:\\ 
\indent count\_helper is\_some\_true (flip\_vote p)  l =\\
\indent count\_helper is\_some\_false p l.
}
\subsection{Monotonicity}
To show that the simple majority function is monotone, I proved a number of lemmas that specify  how the number of each type of preference changes if we update a certain voter's preference for each possible case. Here I present one of them, which states that changing a \texttt{Some false} vote to a \texttt{None} decreases the \texttt{Some false} vote frequency by one:

\centercode{Lemma update\_count\_lemma\_1 v p l \`{}\{NoDup l\}: \\
\indent p v = Some false → In v l →\\
\indent count\_helper is\_some\_false p l =\\
\indent \indent 1 + count\_helper is\_some\_false (update v None p) l.
}

There is also the case where if we are considering a subset of voters, and we update someone not in the subset, this does not affect the frequency of each type of preference within the original subset:

\centercode{Lemma upgrade\_not\_in\_list f p v cand l: \\
$\thicksim$In v l → count\_helper f p l = count\_helper f (update v cand p) l.
}
\section{Only if direction\label{onlyif}}

For the only if direction of the theorem, we prove that for any social choice function, denoted as \texttt{scf}, that is anonymous, neutral, and monotone, it is equivalent to the simple majority function.

Given our goal \texttt{scf = majority\_election}, by the axiom of functional extensionality, it suffices to prove that for any list of preferences, denoted as \texttt{p}, it is the case that \texttt{scf p = majority\_election p}.  Subsequently, we perform  case analysis of all the possible outcomes of \texttt{scf p} and \texttt{majority\_election p}. For the three cases where they match, the statement follows trivially. Thus, it suffices to show that if \texttt{scf p $\neq$ majority\_election p}, we can achieve a contradiction where we can prove the \texttt{False} proposition. 

There are six cases where \texttt{scf p $\neq$ majority\_election p}:
\begin{center}
\begin{tabular}{ c | c | c} 
 Case number & Preference condition & \texttt{scf p} \\ 
 \hline
 1 & \texttt{count is\_some\_false p < count is\_some\_true p} & \texttt{Some false}   \\
 2 & \texttt{count is\_some\_false p < count is\_some\_true p} & \texttt{None}  \\
 3 & \texttt{count is\_some\_true p < count is\_some\_false p} & \texttt{Some true}  \\
 4 & \texttt{count is\_some\_true p < count is\_some\_false p} & \texttt{None}  \\
 5 & \texttt{count is\_some\_false p = count is\_some\_true p} & \texttt{Some false}  \\
 6 & \texttt{count is\_some\_false p = count is\_some\_true p} & \texttt{Some true}  \\
\end{tabular}
\end{center}

Actually, half of the cases are  redundant. For example if case number $3$ holds, we can reduce it to a variation of case number $1$. To see why this is the case, assume case number $3$ holds, and consider the list of preferences with each element flipped, i.e.\  \texttt{flip\_vote p}, denoted as \texttt{p$'$}. By the two lemmas from subsection~\ref{neutrality}, we have \texttt{count is\_some\_false p$'$ < count is\_some\_true p$'$}. In addition, by the neutrality condition of \texttt{scf}, we have \texttt{scf p$'$ = Some false}. We now  have a variation of case number $1$ where we replace \texttt{p} with \texttt{p$'$}. In other words, we only need to find a contradiction for each of the following three cases:

\begin{center}
\begin{tabular}{ c | c | c} 
 Case number & Preference condition & \texttt{scf p} \\ 
 \hline
 1 & \texttt{count is\_some\_false p<count is\_some\_true p} & \texttt{Some false}   \\
 2 & \texttt{count is\_some\_false p<count is\_some\_true p} & \texttt{None}  \\
 3 & \texttt{count is\_some\_false p=count is\_some\_true p} & \texttt{Some false}  \\
\end{tabular}
\end{center}

For the rest of this section, unless otherwise specified, we use \texttt{a} and \texttt{b} to denote the number of \texttt{Some true} and \texttt{Some false} votes in the list of preferences \texttt{p}, respectively.  

\subsection{Case 1: \texttt{a > b} and \texttt{scf p = Some false}}
Suppose \texttt{a > b} and \texttt{scf p = Some false}. Consider \texttt{flip\_vote p}, the list of all the preferences flipped, denoted as  \texttt{p1}. The following three statements can be proved:

\begin{enumerate}
    \item Using proof by induction, the number of \texttt{Some true} and \texttt{Some false} votes in \texttt{p1} are flipped with respect to that of \texttt{p}, i.e.\ \texttt{count is\_some\_false p1 = a} and \texttt{count is\_some\_true p1 = b}.
    \item By case analysis, the preference of a voter in \texttt{p} is \texttt{None} if and only if their preference in \texttt{p1} is \texttt{None}, i.e.\ \texttt{$\forall$ voter, p voter = None $\leftrightarrow$ p1 voter = None}.
    \item By neutrality of \texttt{scf}, we have \texttt{scf p1 = Some true}.
\end{enumerate}

We then construct the list of preferences \texttt{p2}, which is the same as \texttt{p1} but we update the first \texttt{(a $-$ b)} \texttt{Some false} votes in \texttt{p1} to \texttt{Some true} via the function \texttt{upgrade\_vote\_list}\footnote{The Coq keyword \texttt{Fixpoint} is the same as the keyword \texttt{Definition}, except that it is used specifically for recursive definitions. It also performs additional checks that the defined function is total.}:

\centercode{Fixpoint upgrade\_vote\_list p l:=\\
    \indent match l with \\
     \indent   \indent  | [] => p \\
  \indent\indent      | hd::tl => update hd (Some true) (upgrade\_vote\_list p tl)\\
     \indent   end.
}

The following three statements then follow:

\begin{enumerate}
    \item Using proof by induction, the number of \texttt{Some true} and \texttt{Some false} votes in \texttt{p2} are the same as that of \texttt{p}, i.e.\ \texttt{count is\_some\_true p2 = a} and \texttt{count is\_some\_false p2 = b}.
    \item By showing that \texttt{None} votes are not changed from \texttt{p1} to \texttt{p2}, it is the case that the preference of a voter in \texttt{p} is \texttt{None} if and only if their preference in \texttt{p2} is \texttt{None}, i.e.\ \texttt{$\forall$ voter, p voter = None $\leftrightarrow$ p2 voter = None}.
    \item By monotonicity of \texttt{scf} and the following lemma, we have \texttt{scf p2 = Some true}:
   
  \centercode{Lemma upgrade\_vote\_list\_monotone scf p l:\\
  \hspace*{0.5cm} monotone scf → \\
   \hspace*{0.5cm} scf p = Some true → \\
   \hspace*{0.5cm} scf (upgrade\_vote\_list p l) = Some true.}
\end{enumerate}

Lastly, after defining a function for swapping a list of pairs of preferences, we show that there exists a simple list of swaps that enables us to transform \texttt{p2} to \texttt{p}:

\centercode{Fixpoint swaps p l:=\\
 \indent   match l with \\
 \indent       | [] => p \\
   \indent     | (x,y)::tl => swap x y (swaps p tl)\\
 \indent   end.\\
Lemma same\_true\_num\_implies\_swappable p p':  \\
\indent ($\forall$ x : voter, p x = None $\leftrightarrow$ p' x = None) → \\
\indent count is\_some\_true p = count is\_some\_true p' →\\
\indent $\exists$ l , p = swaps p' l.
}

In fact, this list can be constructed easily: it is the result of zipping the list of voters who voted \texttt{Some true} in \texttt{p} and \texttt{Some false} in \texttt{p2}, and the list of voters who voted the other way round, as highlighted by the following function: 

\centercode{Definition count\_true\_same\_swap\_list\_helper p p' l:= \\
 \indent   let l1:= left\_true\_right\_false p p' l in \\
 \indent  let l2:= left\_false\_right\_true p p' l in \\
 \indent   zip l1 l2.
}

Various properties of this list have to be proved. For example, one has to prove that the two lists are of the same length, so no element is dropped during the \texttt{zip} process: 

\centercode{Lemma count\_true\_difference\_relation p p' l \`{}\{!NoDup l\}:\\ 
 \indent   ($\forall$ x, p x = None $\leftrightarrow$ p' x = None) →\\
    \indent count\_helper is\_some\_true p l = count\_helper is\_some\_true p' l →\\
\indent    length (left\_true\_right\_false p p' l) = \\
\indent \indent length (left\_false\_right\_true p p' l).
}

Lastly, by the anonymity property of \texttt{scf}, we then have \texttt{scf p = Some true}. However, we started with the assumption that \texttt{scf p = Some false}, and thus a contradiction is achieved.
\subsection{Case 2: \texttt{a > b} and \texttt{scf p = None}}

In this case, where \texttt{a > b} and \texttt{scf p = None}, the proof is similar to that of case number 1. However, during the construction of \texttt{p2}, we need a different invariant to prove that \texttt{scf p2 = Some true}. Specifically, we use the following lemma together with the fact that the number of voters to be updated is non-zero (since \texttt{a $-$ b > 0}):

\centercode{Lemma upgrade\_vote\_list\_monotone\_weak scf p l:\\
\indent monotone scf → (scf p = Some true $\lor$ scf p = None) →\\
\indent(scf (upgrade\_vote\_list p l) = Some true $\lor$ \\ 
\indent \indent scf (upgrade\_vote\_list p l) = None).
}

\subsection{Case 3: \texttt{a = b} and \texttt{scf p = Some false}}
In this case, where \texttt{a = b} and \texttt{scf p = Some false}, the entire proof for case number 1 almost works perfectly here for us to achieve a contradiction as well. The main difference is that we need not update any voters to produce \texttt{p2} from \texttt{p1} (as \texttt{a - b = 0}), i.e.\ \texttt{p2 = p1}.
\section{Discussion\label{discussion}}

\subsection{Why is this formalization useful?}
The first obvious answer is that formalizing the proof of a theorem allows us to accept it as fact with more certainty. Humans make mistakes, and it is not uncommon to hear mathematical proofs widely accepted by the community are found to contain minor gaps and inaccuracies afterwards~\cite{paulson}\cite{gouezel}. While it is unlikely for May's theorem to be false,  formalizing it enables us to ensure we do not miss any edge cases in high-level proof sketches. 

This project also enables one to gain a precise understanding of the technique used within the proof. Defining and proving properties of each step rigorously allows the programmer to gain deeper insights into how and why a proof works. In particular, from a more personal perspective,  the first proof I wrote relies on the axiom of excluded middle, which is an axiom from classical logic that is not included in Coq's intuitionistic logic system. Only during a thorough review of the code did I realize that the proof can be rewritten slightly to circumvent using the axiom, and thus allowing my most recent proof to be completely constructive. 

Nonetheless, I believe whether a project is useful or not is sometimes not the best way to justify its value. I mainly pursued working on this project simply because I was interested in it. To quote Benthem Jutting's PhD thesis~\cite{jutting}:

\noindent "\textit{A further motive, for the author, was that the work involved in the project appealed to him.}"

\subsection{Is this theorem not obvious?}
It might be true that most intermediate steps of the proof are not difficult to understand. However this is not equivalent to saying that the theorem is obvious or that the project itself is trivial.

Firstly, as mentioned at the beginning, though most intermediate steps might be easy to comprehend, the proof of those steps might be long and tedious. When reading a pen-and-paper proof, we automatically infer various properties implicitly, but in a proof assistant, everything must be stated and proved explicitly, e.g.\ whether a list has no duplicates, whether an element is contained (or not contained) within a list. 

In addition, without knowing the trick beforehand, it is not exactly straightforward how one can prove May's theorem, especially for the only if direction. Perhaps, it might be more accurate to describe the proof to be elegantly short as opposed to describing the theorem as trivial. 

\subsection{Is this proof correct?\label{correct}}
There are two potential sources of  errors which might lead to the proof being incorrect. Luckily, both  are  unlikely.

Firstly, the Coq proof itself might not be sound, meaning that we can trace the error back to a bug in the kernel of the Coq proof assistant. Coq is based on the Calculus of Inductive Constructions~\cite{CIC}, which is a reliable and well-understood type theory. Coq also  satisfies the de Bruijn criterion~\cite{criterion}, meaning it generates proof terms that can be verified by  an independent and relatively small kernel.  As a powerful verification tool that has been maintained for more than thirty years, one can safely trust and assume the correctness of Coq's kernel without losing much sleep. 

Another source of error might be due to the definitions stating a completely different theorem instead of that of May's. To reduce the likelihood of this mistake,  definitions are written as clearly as possible and are written to reflect the original texts closely. 

There is one part where the Coq proof differs from the original paper. In May's paper, there is actually  a fourth property in the set of sufficient and necessary properties for simple majority voting, called \textbf{decisiveness}, which states the function must be defined and single-valued for all possible lists of preferences. This, however, is exactly the definition for a function in mathematics anyway! As so, many subsequent papers on the theorem omitted this property in the statement~\cite{infinite}, which I also followed suit. 

\subsection{Is the axiom of extensionality necessary?\label{extensionality}}
Recall that the mechanical proof assumes the axiom of extensionality, which is not part of Coq's standard library. It is used because when we assert that two social choice functions are equivalent, we implicitly mean that the functions agree on every possible input. 

One can circumvent the need for the axiom by redefining May's theorem slightly:

\centercode{Theorem mays\_thm2 scf: \\
\indent anonymous scf $\land$ neutral scf $\land$ monotone scf $\leftrightarrow$ \\ \indent $\forall$ p, scf  p = majority\_election p.}

I still stuck with the first definition (see Section~\ref{definitions}), since I think that definition of equality is clearer, as justified by Subsection~\ref{correct}.  
\section{Conclusion and future directions\label{conclusion}}
In this report, I showed how May's theorem is formalized in the Coq proof assistant. I discussed several aspects of the Coq proof, including the translation from the theorem statement, various lemmas used in the proof, and its correctness.

Since the original publication, various others have extended May's theorem in multiple ways, e.g.\ when there is an infinite number of voters~\cite{infinite}, or when there are more than two candidates~\cite{list}. A possible extension is to formalize those generalized theorems in Coq as well.

Being the first formalization of May's theorem, this Coq proof establishes another step into formalizing fundamental results in social choice theory.  Another extension would be to continue formalizing other interesting theorems in this area, such as the median voter theorem~\cite{black}, the McKelvey-Schofield chaos theorem~\cite{mckelvey}, and Sen's possibility theorem~\cite{sen}. It might also  be worthwhile to develop a library containing voting-related primitives and lemmas for formalizing similar theorems.


\bibliographystyle{acm}
\bibliography{main}

\end{document}